\documentclass[a4paper,12pt,noshowpacs,nofootinbib,amsmath,amssymb,superscriptaddress]{article}
\usepackage[T1]{fontenc}
\usepackage[utf8]{inputenc}
\usepackage{authblk}
\usepackage{graphicx}
\usepackage{amssymb}
\usepackage{amsmath}
\usepackage{amsfonts}
\usepackage{latexsym}
\usepackage{wrapfig}
\usepackage{floatflt}
\usepackage{hyperref}
\usepackage[all]{xy}
\usepackage{slashed}
\usepackage{physics}
\usepackage{accents}
\usepackage{afterpage}
\usepackage{booktabs,siunitx,dcolumn}
\usepackage{color}
\usepackage[dvipsnames]{xcolor}

\usepackage[labelfont=bf]{caption}
\usepackage[numbers,sort&compress]{natbib}
\usepackage{lipsum}
\setcitestyle{square}

\newcommand{\be}{\begin{equation}}
\newcommand{\ee}{\end{equation}}
\newcommand{\ba}{\begin{eqnarray}}
\newcommand{\ea}{\end{eqnarray}}

\def\d{\delta}

\def\x{{\bf x}}

\def\k{{\bf k}}

\def\[{\left[}
\def\]{\right]}
\def\({\left(}
\def\){\right)}
\def\<{\langle}
\def\>{\rangle}

\def\x{\mathbf{x}}

\def\O{\mathcal{O}}
\def\L{\mathcal{L}}

\def\D{\mathcal{D}}

\def\eps{\epsilon}
\def\Mp{M_{\rm p}}

\def\rarr{\rightarrow}

\def\tzeta{\tilde{\zeta}}

\begin{document}

\begin{titlepage}

\pagenumbering{gobble}

\title{
\Large\bf 
Classical Entanglement Structure in the Wavefunction of Inflationary Fluctuations}

\author{Elliot Nelson$^1$}

\author{C. Jess Riedel$^2$}

\affil{\normalsize Perimeter Institute for Theoretical Physics \\ 31 Caroline St. North, Waterloo, ON N2L 2Y5, Canada}

\maketitle

\noindent \begin{center} {\bf Abstract } \end{center}
\begin{quotation} 
{\normalsize
We argue that preferred classical variables emerge from the entanglement structure of a pure quantum state in the form of redundant records: information shared between many subsystems.
Focusing on the early universe, we ask how classical 
metric perturbations emerge from vacuum fluctuations in an inflationary background. 
We show that the squeezing of the quantum state for super-horizon modes, along with minimal gravitational interactions, leads to decoherence and to an exponential number of records of metric fluctuations 
on very large scales, $\lambda/\lambda_{\rm Hubble}>\Delta_\zeta^{-2/3}$, where $\Delta_\zeta\lesssim 10^{-5}$ is the amplitude of 
metric fluctuations. 
This determines a preferred decomposition of the inflationary wavefunction into orthogonal ``branches''
corresponding to classical metric perturbations,
which defines an 
inflationary entropy production rate
and accounts for the emergence of 
stochastic, inhomogeneous spacetime geometry. 
}
\end{quotation}

\vfill
\flushleft{\small $^1$enelson@pitp.ca}
\flushleft{\small $^2$jessriedel@gmail.com}
\flushleft{Essay written for the Gravity Research Foundation 2017 Awards for Essays on Gravitation.\\ Submitted  \today}
\end{titlepage}
\eject

\setcounter{page}{2}

\pagenumbering{arabic}

The macroscopic world features many distinguished ``classical'' variables, such as the strength of an electric field, the position of a planet, or a spacetime metric configuration. 
At the same time, it is hoped that Nature can be described as a quantum state which evolves unitarily and \textit{a priori} does not distinguish any preferred variables in the Hilbert space.
Primordial cosmology, in particular, describes the universe as quantum in origin, and forces us to ask how the classical world emerges in the absence of an external environment or measurement.
The preferred classical variables for a 
quantum system should in principle be determined algorithmically, given the system's Hamiltonian 
and a decomposition of Hilbert space into subsystems (motivated, or fixed, by the Hamiltonian).
These observables would define a decomposition of the global pure state into simultaneous eigenstates: $\Psi=\sum_i\Psi_i$. Each orthogonal state is a \emph{branch} of the wavefunction, corresponding to distinct stochastic outcomes (indexed collectively by $i$) for the classical variables.


An essential and intuitive feature of the classical world is \textit{redundancy of recorded information} in many environmental degrees of freedom \cite{qdarwin,qdarwin2,zurek1994decoherence}. 
The key idea is that 
the effective irreversibility of apparent wavefunction collapse is guaranteed by amplification processes -- whether natural or, occasionally, in laboratories -- that create many redundant correlations widely spread over space.  
For example, the position of a baseball 
is recorded by the electromagnetic field in many different regions of space, hence its visibility from many directions.
When a Stern-Gerlach experiment records the spin of a particle, it amplifies this information into many subsystems of the measuring device and environment, possibly including an experimentalist's brain.
Quantum measurement in this non-anthropocentric sense -- decoherence with 
amplification of recorded information to macroscopic scales -- is ubiquitous in Nature and underlies classical dynamics.
Ultimately, recorded information is encoded in the \textit{entanglement structure of the quantum state} in the form of long-range correlations between many subsystems, relative to a particular (tensor) decomposition of the Hilbert space into distinct parts \cite{riedel2017classical}.


Candidate theories of quantum gravity generally support quantum superpositions of arbitrary classical spacetimes, raising the question of how configurations of the metric emerge as preferred classical variables.
In this essay, we address a part of this problem, by fixing a classical background spacetime and showing that 
the entanglement structure of the wavefunction of metric fluctuations determines classical gravitational fields as preferred observables.
We consider an inflationary background: approximate de Sitter space with a slowly changing Hubble rate. This is a unique and important system for emergent classicality because
(i) inflation is the leading paradigm for both the early and late $\Lambda$CDM universe; 
(ii) no appeal can be made to external environments or observers despite enormous distances and timescales;
(iii) gravity, unusually, plays a key role in the quantum-classical transition; and
(iv) the model is only weakly nonlinear and is computationally tractable.



\textit{Inflationary Fluctuations: Review.}
We take the most straightforward and conservative approach:
Working in an approximately de Sitter background with scale factor $a(t)=e^{Ht}$ with slowly varying Hubble rate $H$, we treat general relativity (GR), expanded perturbatively around the background, as a low-energy effective (quantum) field theory.
Breaking the time translation invariance of de Sitter space so that the Hubble parameter can change, and inflation eventually end, brings to life a dynamical scalar mode, along with tensor (graviton) modes $\gamma_{ij}$.
Using the ADM formalism and choosing spatial hypersurfaces with uniform matter density \cite{Maldacena:2002vr}, the scalar mode is described as curvature fluctuations $\zeta$ in the spatial metric, 
\be\label{zeta_def}
g_{ij}(\x) = a^2(t)[(1+2\zeta(\x))\delta_{ij}+\gamma_{ij}(\x)]
\ee
at linear order.
The slow variation of the Hubble rate, $dH/dt\equiv-\eps H^2$ with $\eps\ll1$, leads to approximately scale-invariant spectra of scalar and tensor perturbations on super-Hubble scales: 
\be\label{two_point}
\Delta_\zeta^2\equiv\frac{k^3}{2\pi^2}\<\zeta_\k\zeta_{-\k}\>|_{t\rarr\infty} \approx \frac{1}{(2\pi)^2}\frac{H^2}{2\eps\Mp^2},
\hspace{0.7cm}
\Delta_\gamma^2\equiv\frac{k^3}{2\pi^2}\<\gamma^s_\k\gamma^s_{-\k}\>|_{t\rarr\infty} \approx \frac{1}{\pi^2}\frac{H^2}{\Mp^2},
\ee
for either tensor polarization $s=+,\times$.\footnote{We omit a momentum-conserving delta function, $(2\pi)^3\d^3(\k+\k')$, and set $\k'=-\k$.} ($\Mp$ is the reduced Planck mass.)

The quantum state of the fluctuations can be written as a wave functional $\Psi[\zeta(\x),\gamma_{ij}(\x)]$, which has the asymptotic late-time behavior 
\be\label{WF}
\Psi[\zeta,\gamma](t) \rarr e^{iS[\zeta,\gamma](t)}|\Psi[\zeta,\gamma]|
\ee
for the fields on super-Hubble scales.  The action $S$ 
is integrated to time $t$ and grows exponentially as $a(t)$. While post-inflationary observables such as those in Eq. \eqref{two_point} are determined by the (asymptotically constant) amplitude $|\Psi|$, the formation of redundant records will be captured in the phase information, and can be described with correlation functions depending on the conjugate momenta (or field velocities) as well as the fields.

\textit{Gravitational Interactions.}
Combining the quadratic action for $\zeta$, $\L_{\rm free}$, with the leading cubic self-interactions $\L_{\rm int}$ obtained by perturbatively expanding the Einstein-Hilbert action, the Lagrangian in the presence of a long-wavelength background field $\zeta_L$ can be written in the simple form\footnote{Throughout, overdots denote derivatives with respect to cosmic time $t$.} \cite{Maldacena:2002vr,Burrage:2011hd,Creminelli:2011rh,Creminelli:2012ed}
\be\label{L_shift_zeta}
\L_{\rm free}+\L_{\rm int} = \eps\Mp^2 \[ a^3 \(1+3\zeta_L\) \dot{\zeta}^2 - a(1+\zeta_L)(\nabla\zeta)^2 \] + ...,
\ee
where the dots include subdominant derivative interactions and $\O(\eps^2)$ terms.
The effect of the cubic interactions is to replace the scale factor $a(t)$ with a spatially varying scale factor, $a(t)[1+\zeta_L(\x,t)]$.

\begin{figure}[t!]
\hspace{1cm}
\includegraphics[scale=0.52]{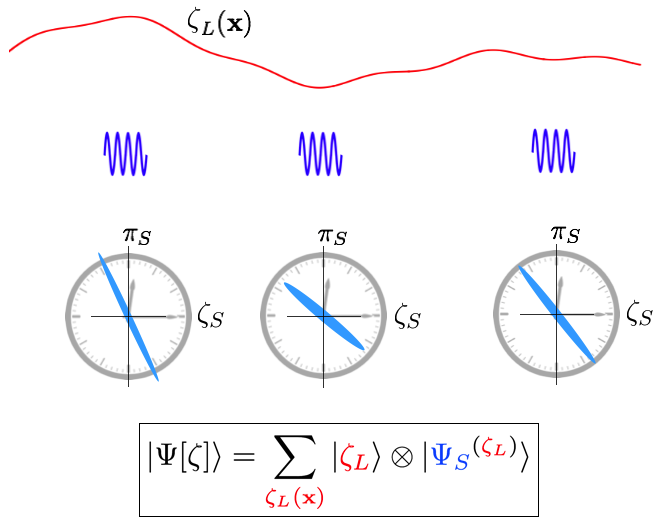}
\caption{A schematic depiction of redundant records of long-wavelength fluctuations during inflation, due to gravitational coupling of short- and long-wavelength modes. The squeezed quantum states for super-horizon modes $\zeta_S$ (localized in different spatial regions), depicted in phase space, are highly time-dependent.
They respond as clocks to the spatially varying ``time'' set by the local background field $\zeta_L$ from much longer modes.
This recording of long-wavelength information appears in the quantum state $|\Psi[\zeta]\rangle$, which can be decomposed into ``branches'' corresponding to long-wavelength field configurations $\zeta_L$, with the conditional state $|\Psi_S^{(\zeta_L)}\>$ for the remaining shorter modes containing many records of $\zeta_L$.}
\label{fig:clocks}
\end{figure}

\textit{Decoherence and Redundant Records.}
It is well known that as a mode redshifts to super-Hubble scales, its quantum state becomes very highly squeezed (appearing in phase space as a thin ellipse), 
with the amount of squeezing determined by the scale factor \cite{Albrecht:1992kf,Polarski:1995jg}.
Since, as we just saw, a long-wavelength background $\zeta_L$ perturbatively shifts the scale factor, it shifts the amount of squeezing for shorter-wavelength modes.
Even a tiny change in squeezing displaces the state to an orthogonal state, so the short modes become extremely sensitive to the background $\zeta_L$, recording it like clocks (Figure \ref{fig:clocks}).
These records are captured in the entanglement structure of the quantum state $|\Psi[\zeta]\>$.\footnote{We suppress the tensor argument $\gamma_{ij}$ when discussing scalar modes.} This is illustrated in Figure \ref{fig:clocks}, and can be seen from the appearance of the interactions, Eq. \eqref{L_shift_zeta}, in the phase oscillations in the wavefunction, Eq. \eqref{WF}.

The number of short-wavelength modes that have crossed the horizon,
and therefore also the number of ``clocks'' that can store a record of the local background, scales with the total volume, $\propto a^3(t)$.
Thus, the number or \textit{redundancy} $R_\zeta$ of records of the average background $\zeta_L$ in the shorter-wavelength environment grows rapidly. This can be computed as a function of the \textit{precision} of the records $\Delta\zeta_L$, and is \cite{Nelson:2016kjm,inflation_branching}
\be\label{R}
R_\zeta(\Delta\zeta_L,aHL) = \O(1)\cdot\frac{\Delta\zeta_L^2}{\<\zeta_L^2\>} \frac{H^2}{2\eps\Mp^2}(aHL)^3,
\ee
where $L$ is the comoving length of the region covered by $\zeta_L$, and $aHL$ is the physical wavelength in Hubble units.
The redundancy first becomes $\gtrsim1$ for very imprecise records, with $\Delta\zeta_L^2$ not much smaller than the range of possible values $\<\zeta_L^2\>$. This occurs when a mode redshifts to the physical scale
\be
\lambda^{(\zeta)}_{\rm classical} \equiv (aL)_{\rm classical} \equiv \Delta_\zeta^{-2/3}\cdot H^{-1},
\ee
at which it begins to decohere.
For $\Delta_\zeta^2\sim 10^{-10}$ fixed to the amplitude of temperature fluctuations in the cosmic microwave background, this is $\O(1000)$ Hubble distances.
After this point, the redundancy grows as the volume, $\sim a^3$, and the squared coherence length (the spread in field values for which quantum coherence remains, which is the precision at which $R_\zeta\lesssim1$)
decays as $1/a^3$ \cite{Nelson:2016kjm,Burgess:2014eoa}.






While we have used the comoving gauge, the production of records of the long-wavelength field can also be described in terms of density fluctuations in the spatially flat gauge.  In that gauge, a background fluctuation shifts the local energy density or Hubble rate, which distorts shorter super-Hubble modes in a similar way.


\textit{Primordial Gravitational Waves.} While we've focused on scalar fluctuations, tensor modes behave similarly. They couple to scalar modes at leading order via the interaction
 \cite{Maldacena:2002vr}
\be\label{L_gzz}
\L^{(\gamma\zeta\zeta)}_{\rm int} = \eps \Mp^2 a(t) \gamma_{ij}\nabla_i\zeta\nabla_j\zeta.
\ee
We see that a tensor background $\gamma_{L,ij}$ effectively shifts the gradient energy of a scalar mode, $k^2\rarr k^2-\gamma_{ij,L} k^i k^j$, which also distorts its quantum state as it becomes squeezed.
This allows super-Hubble scalar modes to be a bath of detectors of the longer-wavelength tensor background. Because tensor modes have a smaller amplitude, $\Delta_\gamma^2\sim H^2/\Mp^2$, they are harder to detect, and decohere only after reaching a larger wavelength,
\be
\lambda^{(\gamma)}_{\rm classical} = \Delta_\gamma^{-2/3}
\cdot H^{-1} \simeq \eps^{-1/3}\cdot\lambda^{(\zeta)}_{\rm classical}.
\ee

\textit{Entropy.} As more modes redshift to the scales $\lambda_{\rm classical}^{(\zeta),(\gamma)}$ and their stochastic amplitudes are redundantly recorded, the entropy of the inflationary universe grows.
Adding the contributions of all modes which have decohered \cite{Kiefer:2006je}, the rate of entropy production per unit volume is \cite{inflation_branching}
\be
\frac{ds_\zeta}{dt} \sim H^4\Delta_\zeta^2 \sim \frac{H^6}{\eps\Mp^2}, \ \ \ \ \ \ \frac{ds_\gamma}{dt} \sim H^4\Delta_\gamma^2 \sim \frac{H^6}{\Mp^2},
\ee
for scalar and tensor modes, respectively. That is, entropy production is controlled by the expansion rate $H$ along with the interaction strength, which is just the amplitude of scalar or tensor fluctuations.
Despite their weakness, the interactions produce an entropy proportional to the growing volume. For 60 $e$-folds of inflation and $\Delta_\zeta^2\sim10^{-10}$, for example, $S_\zeta\sim 10^{69}$. This describes the randomness in the post-inflationary primordial fluctuations.

\textit{Conclusion.}
We studied general relativity in an inflationary background 
with minimal matter (a dynamical scalar mode),
and saw that gravity plays a key role in the emergence of a classical inhomogeneous spacetime: 
(I) By coupling a constant energy density source to spacetime geometry to drive inflation, gravity makes possible a \textit{super-horizon regime} in which modes of light quantum fields become highly squeezed.
(II) By coupling metric fluctuations on different scales, gravity allows squeezed modes to collect exponentially many records of the long-wavelength field.

Redundant information is a robust and general principle for identifying emergent classicality and branch structure of a quantum state from first principles \cite{riedel2017classical}. It can be computed explicitly for inflationary metric fluctuations, 
leading to a superposition of states corresponding to different classical long-wavelength fields (Figure \ref{fig:clocks}).
The preceding results pave the way for generalizations beyond fixed classical backgrounds, 
where significant new conceptual issues lie.
Future work will also investigate how the redundant recording of classical information in different Fourier modes relates to the redundant recording of information in spatially disjoint regions which accompanies more familiar instances of classicality. 






\bibliographystyle{acm_unsrt}
\setlength{\bibsep}{6.5pt}
\bibliography{deco_grav}

\end{document}